\documentclass[aps,prl,reprint,groupedaddress,twocolumn]{revtex4-1}
\usepackage{graphicx}
\usepackage{mathrsfs}

\setcitestyle{super}

\newcommand{\ket}[1]{|#1\rangle}
\newcommand{\be}{\begin{eqnarray}}
\newcommand{\ee}{\end{eqnarray}}
\begin{document}


\title{Generalized thermodynamics of an autonomous micro-engine}


\author{Lukas Gilz}
\email[]{lgilz@rhrk.uni-kl.de}
\author{Eike P. Thesing}
\author{James R. Anglin}

\affiliation{\mbox{State Research Center OPTIMAS} and \mbox{Fachbereich Physik,} \mbox{Technische Univerit\"at Kaiserslautern}}

\pacs{}
\date{\today}

\begin{abstract}
We analyze an autonomous micro-engine as a closed quantum mechanical system, including the work it performs and the fuel it consumes. Our model system shows by example that it is possible to transfer energy steadily and spontaneously between fast and slow degrees of freedom, in analogy to the way combustion engines convert chemical energy into work. Having shown this possibility, we observe close analogies between the closed-system quantum dynamics of our micro-engine and the First and Second Law of Thermodynamics. From these analogies we deduce a generalized formulation of thermodynamics that remains valid on the micro-scale.Ê
\end{abstract}

\maketitle

Recent technological advances have spurred a renaissance in basic research on the microscopic foundations of thermodynamics \cite{therm0,therm1,therm2,cheneau_light-cone-like_2012,rigol_thermalization_2008}, which was originally developed as a pragmatic description of engines\cite{Carnot1824}. We return to that starting point from a modern perspective but depart from previous concepts of quantum engines\cite{hanggi_artificial_2009,fialko_isolated_2012,astumian_thermodynamics_1997,kieu_second_2004,cohen_2006,kim_quantum_2011,lotze_2012,test,berna,fennimore,tierney,Vale,Rayment,Myong} by remaining strictly within \textit{closed-system} quantum mechanics. Since it is always in principle possible to incorporate all relevant degrees of freedom within the system being studied, rather than taking an open-system picture, it must in principle be possible to describe any combustion engine with a Hamiltonian. What is the simplest Hamiltonian that can describe an engine? The question is fundamental as well as pragmatic, since the unitary evolution that this Hamiltonian generates must represent a microscopic limit of thermodynamics. In this paper we identify a Hamiltonian that describes a minimal \textit{autonomous micro-engine}, and from its behaviour propose a generalized thermodynamics that should apply to any dynamical system, regardless of size.

By an ``autonomous micro-engine'' we mean one that retains two important features of larger engines. As an engine, it converts densely stored energy (`fuel') into work. Being autonomous, it is fully self-contained, requiring no power or control from outside. By incorporating these two features, we depart fundamentally from previous discussions of quantum engines, which have provided important insights but have always included either non-Hamiltonian heat reservoirs or parameters with imposed time dependence, and hence modeled key aspects of engine operation phenomenologically, rather than explaining them from first principles. We consider instead a particular \textit{closed} system with a \textit{time-independent, Hermitian} Hamiltonian of the form
\begin{eqnarray}\label{}
	\hat{H} = \hat{H}_{W}+\hat{H}_{F}+\hat{H}_{\mathrm{ME}}\;.\label{eq:engineH}
\end{eqnarray}
Here the first term $\hat{H}_W$ refers to the work that the engine performs, $H_F$ describes the fuel it consumes, and $H_{ME}$ defines the engine's actual mechanism itself. A schematic view of our particular system is shown in Figure 1. Any engine can in principle be described in this way, if all relevant degrees of freedom are properly incorporated in the system, but it has been assumed that this approach must be prohibitively complex. We will show that it need not be so, though our $\hat{H}_{\mathrm{ME}}$ will be highly nonlinear.

As Carnot pointed out, an engine can be considered a device for lifting a weight, because all other forms of work are equivalent to this \cite{Carnot1824}. We therefore take \begin{eqnarray}\label{}
	\hat{H}_{W} = \frac{M\hat{v}^2}{2}+Mg\hat{z}\;,\label{eq:engineHW}
\end{eqnarray}
referring to a quantum particle with mass $M$, height $\hat{z}$ and upward velocity $\hat{v}$, subject to a constant gravitational field with acceleration $g$. (We refer to velocity rather than momentum in order to reserve the letter 'p' for probabilities, but we maintain the canonical relation $[\hat{z},M\hat{v}]=i\hbar$.) Our micro-engine will lift this weight.

To do so the engine must extract energy from some fuel source, represented by $\hat{H}_F$ in (\ref{eq:engineH}). Since energy is proportional to frequency in quantum mechanics, engines delivering high power for their size must harness high frequency degrees of freedom. A conventional combustion engine, for example, may lift a weight by turning a crank at a few thousand revolutions per minute, but to do so it extracts chemical energy from quantum transitions with frequencies on the order of $10^{14}$ Hz.  It is ultimately due to the enormity of this frequency ratio that engines deliver practical power from portable fuel: the invention of combustion engines, therefore was a discovery of how to transfer energy across large frequency ratios, from fast to slow degrees of freedom. For the fuel of our minimal model, then, we consider a system of bosons that may each occupy either a ground or an excited state $|\mp\rangle$, with a high transition frequency $\Omega$:
\begin{eqnarray}\label{}
	\hat{H}_{F} = \frac{\hbar \Omega}{2}\left(\hat{a}^{\dagger}_+\hat{a}_+ -\hat{a}_-^{\dagger} \hat{a}_-\right)\;.\label{eq:engineHF}
\end{eqnarray}
Here the canonical second-quantized operators $\hat{a}_{\pm}^{\dagger}$ ($\hat{a}_{\pm}$) create (destroy) bosons in either state.

\begin{figure}
\centering
\includegraphics[width=0.3\textwidth]{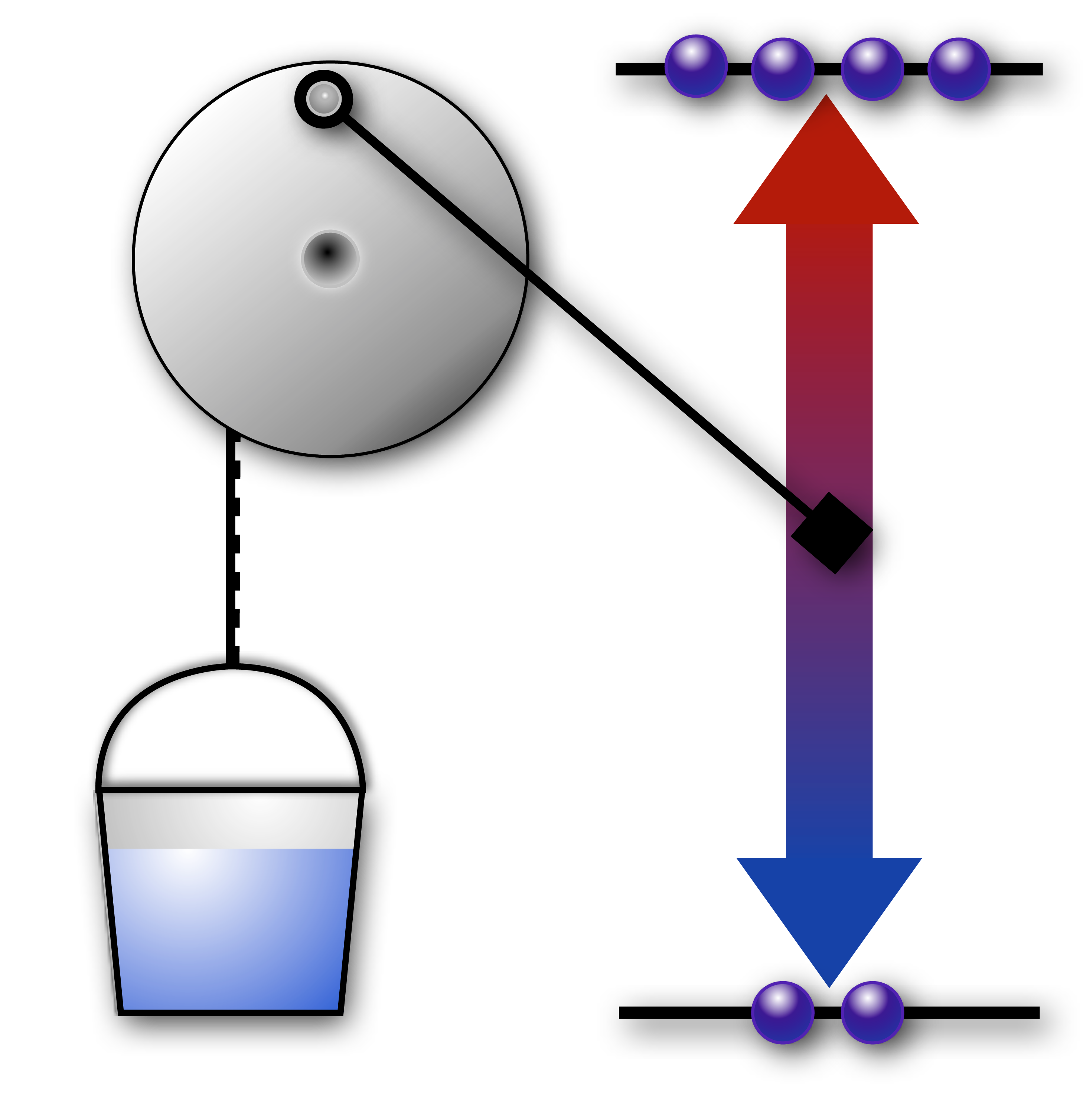}
\caption{Schematic design of the micro-engine (\ref{eq:engineH}). Conserved particles populate a two-mode system with a large frequency gap, and transitions depend on the height of a weight.}
\end{figure}

The role of the remaining term $\hat{H}_{\mathrm{ME}}$ is now simple but difficult: to transfer energy steadily from $\hat{H}_{\mathrm{F}}$ to $\hat{H}_{\mathrm{W}}$. This is difficult because the large frequency gap $\Omega$, which makes the system an engine, is a dynamical challenge. For the natural forms of interaction that are normally considered in physics, direct energy exchange between fast and slow degrees of freedom is never steady, but only a rapid, low-amplitude oscillation, whose neglect is the basis of workhorse approximations in most physical disciplines (adiabatic elimination in atomic physics and quantum optics \cite{Born_Beweis_1928, Grynberg_Introduction_2010}, the Born-Oppenheimer approximation in molecular physics and chemistry \cite{Born_Zur_1927}, and effective Lagrangians for low-frequency modes in high energy \cite{Weinberg} and condensed matter physics\cite{Shankar_Renormalization_1994,Berges_Non_2002}). The principal exception to the rule that fast and slow modes decouple is precisely thermodynamics. In systems with very many degrees of freedom, energy can pass from fast to slow modes by cascading through a long sequence of modes with intermediate frequencies, never crossing any large frequency gaps. The collisions of hot gas particles with a piston, for example, are nearly elastic, and so their high quantum frequencies change only slightly in each interaction. The rapid evolution of such high-frequency degrees of freedom constitutes `the kind of motion we call heat' \cite{Clausius}, which can be converted into low-frequency work over many small steps. The option of cascading across many small frequency steps is not available, however, in a system that has very few degrees of freedom. Is an autonomous micro-engine then possible?

It is, if certain highly nonlinear interactions can be engineered artificially. A sufficient example is this:
\be  
\hat{H}_{\mathrm{ME}}&=& - \frac{\hbar\gamma}{2}\left(e^{ik\hat{z}}\hat{a}_-^{\dagger}\hat{a}_++e^{-ik\hat{z}}\hat{a}_+^{\dagger}\hat{a}_-\right)\label{eq:engineHint}\;,
\ee
where $\gamma\ll \Omega$ and $k$ are constants with dimensions frequency and inverse length, respectively. $\hat{H}_{\mathrm{ME}}$ transfers bosons between the two states, but no term in $\hat{H}$ makes any net change in their total number, which we may thus take as a constant $N$. Note that $\hat{H}_{\mathrm{ME}}$ adds no new degrees of freedom beyond the weight and fuel, and in this sense our micro-engine model is obviously minimal: one degree of freedom each for fuel and work, and none for the mechanism itself.

\begin{figure}
\centering
\includegraphics[width=0.47\textwidth]{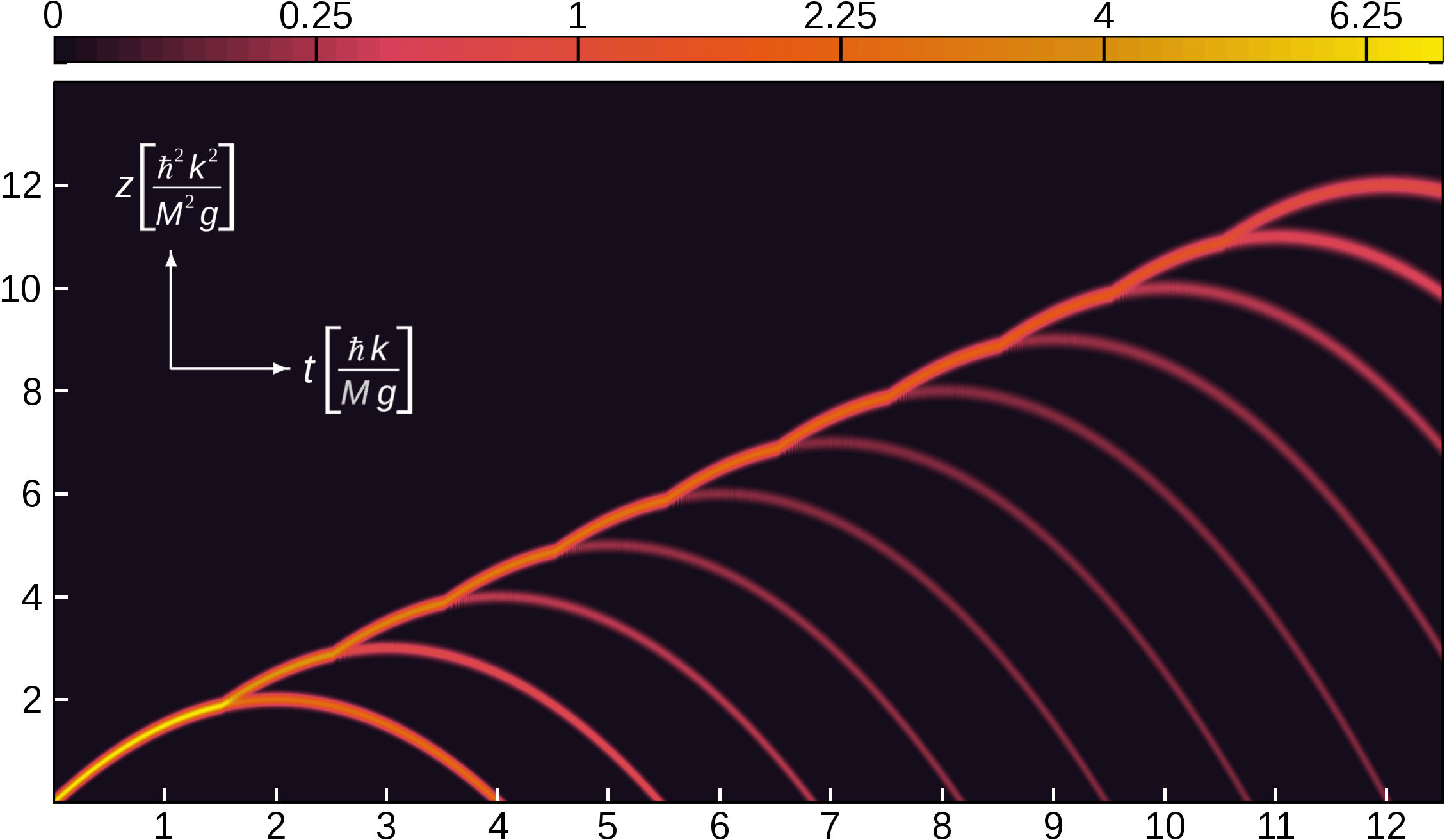}
\caption{Probability density in $\left[1/\frac{\hbar^2 k^2}{M^2 g}\right]$ of the weight's height as a function of time, with nonlinear color scaling to enhance detail. $N=10$ bosons initially occupy the $+$-mode and the weight's initial state is a Gaussian wavepacket, with mean velocity $\Omega/k+\hbar k/M$ and velocity width $(\hbar k/M)/128$. Parameters are $\Omega=\hbar k^{2}/M$, $\frac{\hbar k^2}{M\gamma}=160$ and $\frac{kg}{\gamma^2}=16$.}
\label{fig:EngineWork}
\end{figure}

\begin{figure}
\centering
\includegraphics[width=0.47\textwidth]{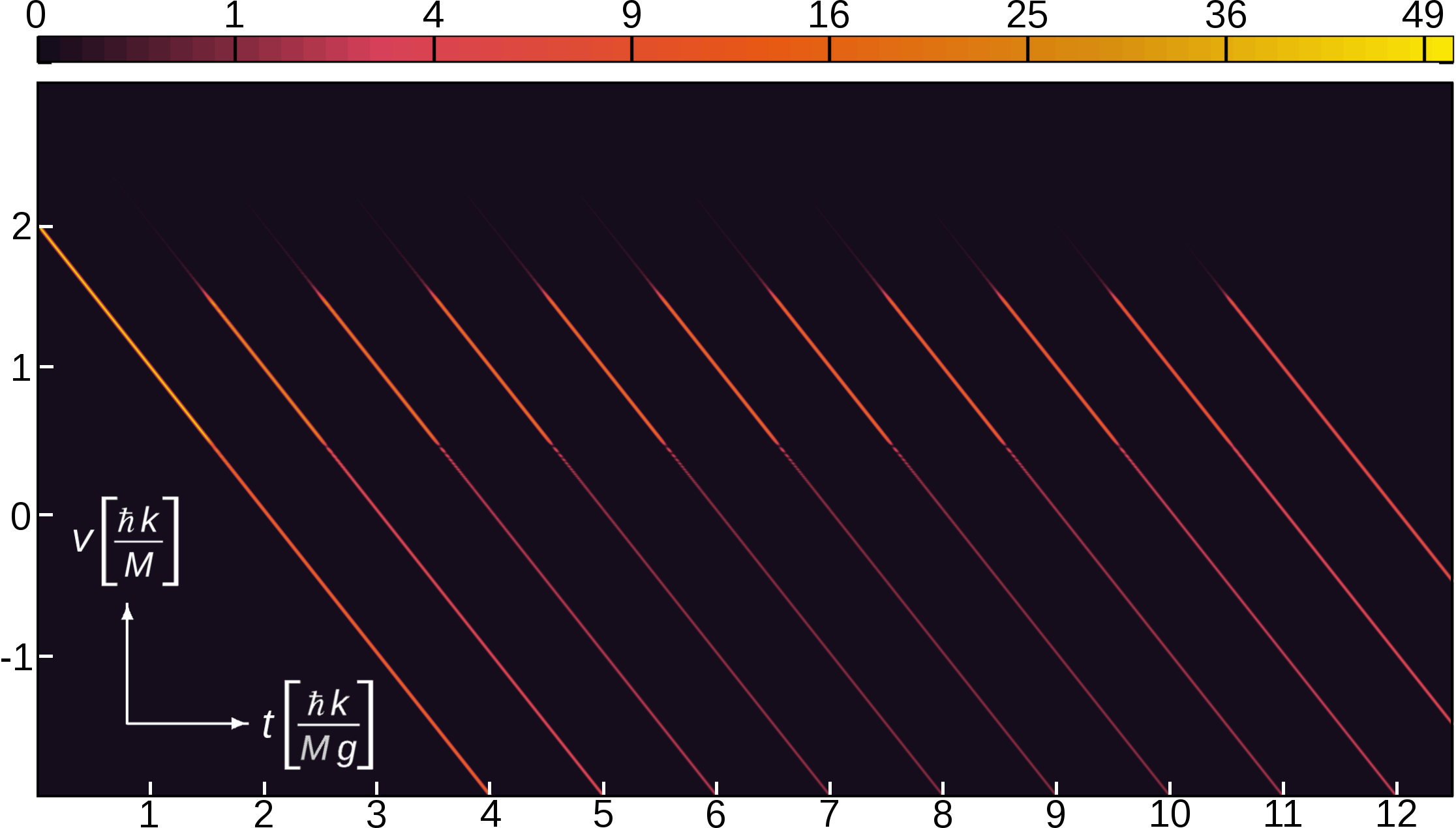}
\caption{The probability distribution $p(v,t)$ of the weight's velocity in $\left[1/\frac{\hbar k}{M}\right]$, incoherently summed over boson distributions, for the same evolution as shown in Figure~\ref{fig:EngineWork}.}
\label{fig:VelDist}
\end{figure}

To see that $\hat{H}$ really does perform as an engine, we evolve the system's quantum state $|\psi(t)\rangle$ by numerically solving the Schr\"odinger equation with the Hamiltonian (\ref{eq:engineH}-\ref{eq:engineHint}), for instructive values of constant parameters. We express the solution in Figure \ref{fig:EngineWork} by plotting the probability distribution of the weight's height $\hat{z}$ summed over boson states $m$
\be
p(z,t)= \sum_{m=0}^N|\langle m|_+\langle N-m|_-\langle z|\psi(t)\rangle|^2
\ee
as a function of time. Some of the engine's features become more apparent in the weight's velocity distribution $p(v,t)$, obtained by the usual Fourier transform from the position representation and shown in Figure \ref{fig:VelDist}. We start from an initial state $\ket{\psi(0)}$ in which all $N$ bosons occupy the $+$-mode, and the weight's state is a Gaussian wave packet, narrow in $v$ around the mean velocity $\Omega/k+\hbar k/M$. Initially launched upwards, the rising weight decelerates under gravity until it has slowed to the critical velocity $v_{c}=\Omega/k-\hbar k/(2M)$. At this velocity, the energy $\hbar\Omega$ of the bosonic transition matches the kinetic energy change if the weight velocity should increase by $\hbar k/M$.  Since such a velocity jump is precisely the operation of $e^{ik\hat{z}}$, $\hat{H}_{\mathrm{ME}}$ becomes a resonant coupling, and the result is that the probability distribution forks. With some probability, the weight simply continues to decelerate under gravity, following the familiar parabolic trajectory. With greater probability (for these parameters), however, the weight makes a quantum jump of $\hbar k/M$ in upward velocity, visible as a kink in the spatial probability distribution of Figure~\ref{fig:EngineWork}. 

After each jump, gravitational deceleration continues, and when the weight slows again to $v_{c}$, the forking repeats. The probability of the velocity jumping up from $v_{c}$ is even higher for subsequent forkings, and so there is a substantial probability that the weight will continue to rise against gravity, at the average velocity $\Omega/k$, until at most $N$ forkings have occurred. After this point, with whatever probability remains, the weight accelerates downwards under gravity. Since $N$ may be arbitrarily large, however, the weight can in principle be lifted to arbitrary height. Examining the full quantum state $\ket{\psi(t)}$, we can confirm that the work done on the weight is exactly matched by the energy lost from the boson system, and that each quantum jump in the weight's velocity occurs with a transition of one boson from the $+$-state to the  $-$-state. The micro-engine consumes its fuel quantum by quantum.

From $p(v,t)$ in Figure \ref{fig:VelDist} we can see that the micro-engine induces true quantum jumps in the weight velocity, and not just short bursts of high acceleration:  the instant at which the velocity jump occurs is probabilistically spread over a short interval, but there is never any probability, during this interval, for the weight velocity to take any intermediate values. It is expected that a quantum system that is slowly driven, as our boson system is by our weight, will exhibit such energy jumps; see, for example, Figure~8 of Ref.~\cite{cohen_2006}. Because our entire system is closed, we see here the corresponding back-action jumps of the slow weight. 

The evolution represented in Figure \ref{fig:EngineWork} is a quantum analog to the familiar situation of driving a car uphill at a nearly steady speed. The steadiness of this speed is what is meant by the weight being `slow'. Demonstrating that this behavior occurs is the non-trivial crux of our paper: autonomous micro-engines are possible. Since conservation of total energy is already guaranteed in closed-system mechanics, our single example proves the general statement: \textit{While total energy is conserved, it is possible to transfer energy between fast and slow degrees of freedom.} This statement is an obvious analog of the First Law of Thermodynamics. But is it only an analog? An established perspective on thermodynamics, dating back to Clausius \cite{Clausius}, makes time scales the essential difference between heat and work in any case. We therefore propose the italicized statement above as the generalized formulation of the First Law. It reduces via Clausius to the standard formulation in macroscopic systems, but it applies to microscopic systems as well.

Along with the possibility of raising the weight, the multiple probability branches of Figure \ref{fig:VelDist} show that each velocity jump may also fail to occur, leading to an average engine efficiency lower than one. This observation will lead to our proposed generalization of the Second Law of Thermodynamics. As the detailed analysis in the supplementary material shows, we can understand the micro-engine's inefficiency analytically in the simplest limit, where $N\gamma \ll\hbar k^{2}/M$ as in Figure~\ref{fig:EngineWork}, by applying Landau-Zener post-adiabatic theory \cite{Zener, Landau}. This yields the approximate probability of the lower branch in the $n$-th bifurcation
\be\label{Pn}
\mathcal{P}_{n} = \exp\left[-\frac{\pi\gamma^{2}}{2kg}n(N+1-n)\right] \;.
\ee
In practical terms this is the probability that the micro-engine, having successfully `burned' $(n-1)$ quanta of fuel, spontaneously stalls, as a quantum fluctuation, and fails to do any further work. 

We speak of probabilities because these describe the results of measurements on the weight alone, but the full evolution is quantum mechanically unitary, and the final state of the entire system is a coherent `Schr\"odinger's Cat' superposition of all different population distributions of the bosons, with correspondingly different wave packets for the weight (all at different heights). This is a pure quantum state, but it is one in which the fast and slow degrees of freedom are highly entangled. The only observables that could be sensitive to the quantum coherence between the branches in this superposition would be those that directly couple states with different $n$. The Heisenberg time evolution of such operators, under $\hat{H}$, involves rapid phase factors $e^{\pm i\Omega t}$. Observing quantum interference between the superposed branches of the highly entangled state $\ket{\psi}$ thus requires measurements that can resolve the short time scale $1/\Omega$. Inspired by the micro-engine $\hat{H}_{\mathrm{ME}}$ itself, one might perhaps evade this requirement by including a factor $e^{\pm ik\hat{z}}$ in observables; but this would require spatial resolution on the short scale $1/k$. Observing quantum interference between the Schr\"odinger's Cat branches therefore requires high resolution in either space or time. As far as coarse-grained measurements that lack such resolution are concerned, the actual unitary evolution of the system is indistinguishable from the probabilistic evolution in which the micro-engine has a random chance, given in the limit $N\gamma \ll\hbar k^{2}/M$ by (\ref{Pn}), of spontaneously stalling instead of performing its $n$th quantum of work.

The efficiency of the micro-engine can be high, because $\mathcal{P}_{n}$ can be low; but $\mathcal{P}_{n}$ cannot be zero, so efficiency cannot be perfect. If the system begins in an eigenstate of high-frequency energy, there is always a chance that some of this energy fails to transfer to the low-frequency sector. This limitation must apply to any autonomous micro-engine that is at all like our model, because quantum adiabatic approximations are not exact except in trivial cases. This suggests the general conjecture: \textit{It is impossible for an autonomous system to convert any definite amount of high-frequency energy entirely into low-frequency energy with unit probability.} This statement constitutes a close analog to the Kelvin-Planck formulation of the Second Law of Thermodynamics \cite{Kelvin}, according to which no process can convert heat entirely into work. From the identification of heat and work as high- and low-frequency energy, again according to Clausius \cite{Clausius}, we therefore propose it as a generalization of the Second Law, which shall apply to all dynamical systems, large and small. The generalized Second Law remains probabilistic even in small systems, which evade the ensemble arguments of conventional statistical mechanics, because probabilities arise from the effective low-frequency decoherence of quantum superpositions. 

The inference from micro-engine inefficiency to the Second Law is by no means superficial. As we have noted, the imperfect adiabaticity of micro-engine operation directly implies growing entanglement between fast and slow degrees of freedom. This means the effective loss of information about slow evolution, as that information is transferred into entanglement that is unobservable within the slow sector. This is the same general phenomenon that is thought to explain thermodynamic entropy increase in macroscopic closed systems, whose exact evolution is after all just as unitary as that of small ones. We have here demonstrated this phenomenon explicitly from first principles. It should not be difficult to test our proposals experimentally, since the essential requirement is not to realize our model exactly, but only to achieve similarly modulated coupling between fast and slow degrees of freedom, and prospects for doing this with trapped atoms or ions, for example, are easy to envision. In the longer term, perhaps the heroic ingenuity of the Steam Age will flourish again on the nano-scale, and micro-engineers will craft practically useful micro-engines. Our demonstration that autonomous micro-engines are possible has in any case shown that thermodynamics is not irreducibly macroscopic after all, but may be universal. In particular, our generalized formulation of the Second Law of Thermodynamics is now formulated within a Hamiltonian framework, so that it can be tested with the theoretical and experimental tools of closed-system quantum mechanics.

\section*{Acknowledgements}
JRA gratefully acknowledges valuable discussions with Sandro Wimberger, Herwig Ott and Artur Widera. LG acknowledges funding from the German Federal Excellence Initiative (DFG/GSC 266).

\section{Author Contribution}
LG and JRA developed the model Hamiltonian. All authors discussed the results and prepared the manuscript. 

\section{Author Information}
The authors declare no competing financial interests. Correspondence and requests for materials should be addressed to LG (lgilz@rhrk.uni-kl.de) or JRA (janglin@physik.uni-kl.de)



\newpage
.
\newpage

\section*{Supplementary Information}
\section*{Supplementary Discussion}

\setcounter{figure}{0}
\setcounter{equation}{0}
\renewcommand{\figurename}{Supplementary Figure}

We can understand the micro-engine dynamics by exactly mapping the already simple Schr\"odinger equation generated by (\ref{eq:engineH}) onto an even simpler one. Without loss of generality we take the total quantum state $\ket{\psi}$ to be an eigenstate of $\hat{N}$ with eigenvalue $N$, and represent $|\psi\rangle$ as
\be
\ket{\psi(t)}\!=\! \sum_{m=0}^N \int\! \!dv\,  \psi_m(v,t) \ket{m}\!_+\ket{N\!-\!m}\!_-\ket{v\!-\!m\hbar k/M}\;\;\; \label{eq:ansatz1}
\ee
where $\ket{v}$ are eigenstates of $\hat{v}$ with eigenvalue $v$. By then re-writing
\be\label{ReductionMapping}
\psi_{m}(v,t)=:\varphi_{m}\left(v + g t-\frac{N\hbar k}{2M},t\right)e^{i\frac{(Mv-\frac{N}{2}\hbar k)^{3}}{6M^{2}\hbar g}}
\ee
we eliminate derivatives with respect to $v$ from the Schr\"odinger equation for $\varphi_{m}(v,t)$:
\be\label{ReducedSchroedinger}
i\partial_{t}\varphi_{m}(v,t) &=& \sum_{l=0}^{N}\left[h_{m}(t-\frac{v}{g}+\frac{\Omega}{kg})\,\delta_{lm}-\tilde{h}_{lm}\right]\varphi_{l}\nonumber\\
h_{m}(t)&=& \frac{\hbar k^{2}}{2M}(m-\frac{N}{2})^{2}+kg t (m-\frac{N}{2})\\
\tilde{h}_{lm}&=&\frac{\gamma}{2}\sqrt{m(N-m+1)}\,\delta_{l,m-1}+(l\leftrightarrow m)\;.\nonumber
\ee
The time dependence of $h_{m}(t)$ is strictly an artifact of the exact reduction to a single dynamical degree of freedom via (\ref{ReductionMapping}). The term in $h_{m}$ proportional to $(m-N/2)^{2}$ is likewise induced by the mapping, and can be considered an effective interaction between the bosons mediated by their interaction with the weight. It is not due to any direct interaction between the bosons, since in $\hat{H}$ there is none. The full system of weight plus fuel still evolves under the time-independent $\hat{H}$. Solving Eqn.~(\ref{ReducedSchroedinger}) for $\varphi_{m}$ with $v$ as a fixed parameter, and inserting the solution into (\ref{ReductionMapping}) to obtain $\psi_{m}(v,t)$, yields the solution for the full system. 

Although the mapping (\ref{ReductionMapping}) is only exact because the weight's potential $Mg\hat{z}$ is linear, a similar mapping will be valid as a Born-Oppenheimer approximation for a wide range of potentials $V(\hat{z})$. 

\begin{figure}
\centering
\includegraphics[width=0.47\textwidth]{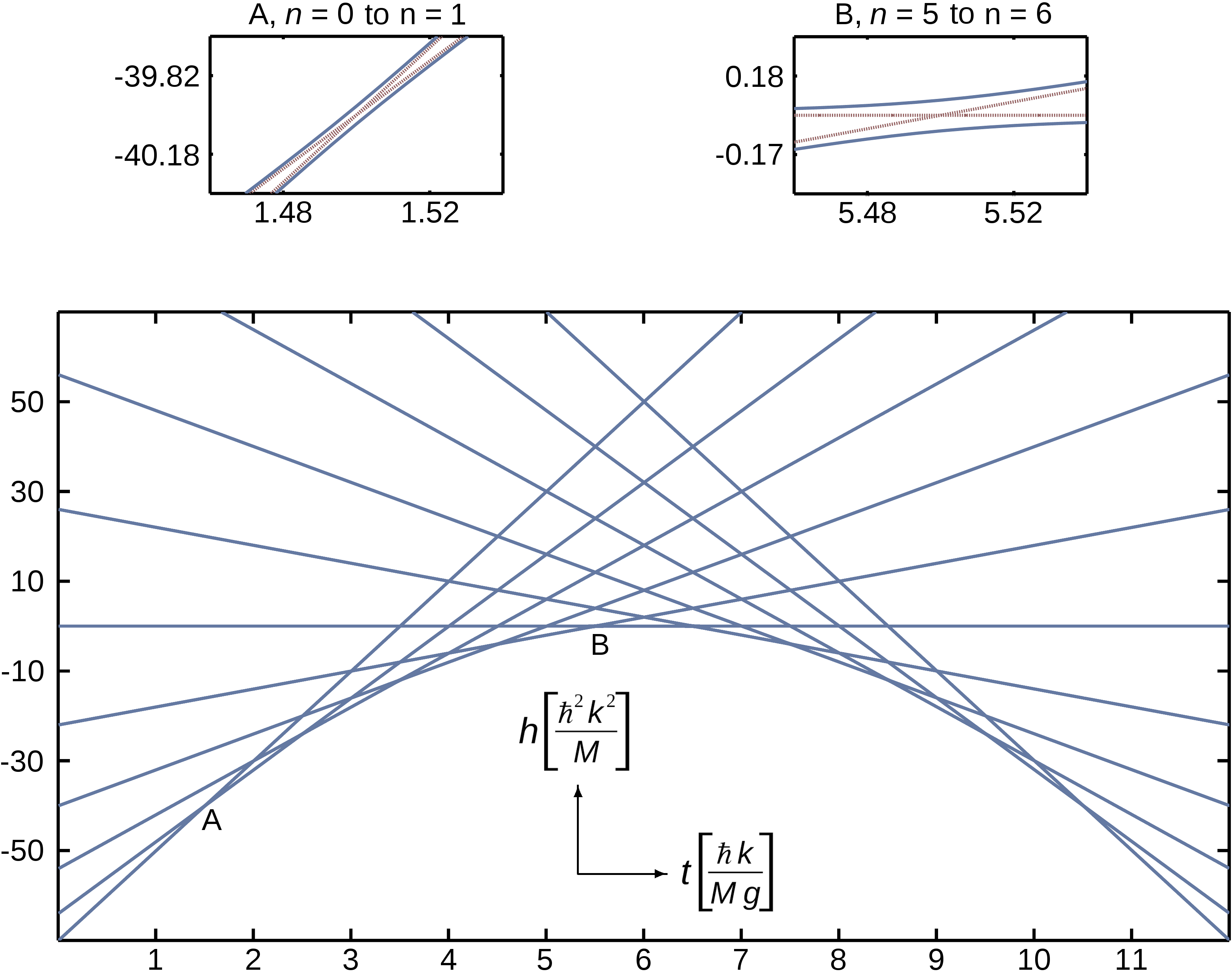}
\caption{Instantaneous energy levels. The main plot shows the set of instantaneous eigenvalues of the full effective Hamiltonian whose matrix elements are $h_{m}(t)\delta_{lm}-\tilde{h}_{lm}$ from (\ref{ReducedSchroedinger}) (blue), as functions of $t$, for $N=10$ particles and all parameters as in Figure~\ref{fig:EngineWork} above. The functions $h_m$ alone (purple, dashed) are also plotted but can only be seen to differ from the blue lines in the small plots above, which show the surroundings of the two crossings marked A and B, in the lowest arc of the web. For larger $N$ the plot remains similar but becomes more dense.}
\label{fig:energylevels}
\end{figure}

We can understand the dynamics described in (\ref{ReducedSchroedinger}) by analyzing its instantaneous energy eigenspectrum, which in the limit where $N\gamma \ll\hbar k^{2}/M$ consists of a series of well separated avoided crossings, as exemplified in Supplementary Figure~\ref{fig:energylevels}. We can solve the general problem by referring to the results obtained independently by Landau and Zener in 1932 \cite{Landau, Zener}. Both considered a two-state Hamiltonian 
\be
\hat{H}=\left(\begin{array}{cc} \epsilon_1 & \epsilon_{12}\\\epsilon_{12} & \epsilon_2\end{array}\right)
\ee
with time-dependent diagonal entries
\be
\frac{2\pi}{h}(\epsilon_1-\epsilon_2)=\alpha t.
\ee
Computing the exact time evolution, Landau and Zener showed that corrections to the adiabatic evolution are negligible, except in the brief episode of the avoided crossing of the instantaneous energy eigenstates. Although brief, this episode may have lasting consequences: the probability that during it the system will jump from one adiabatic state to the other is
\be
P=\exp\left[-\frac{2\pi}{\hbar^2}\frac{\epsilon^2_{12}}{\alpha}	\right]\;.
\ee
In the limit $N\gamma \ll\hbar k^{2}/M$ of our problem, the evolution is adiabatic except very near the avoided level crossings. In particular, therefore, the numbers of bosons in $+$ and $-$ levels remain constant except near those crossings. In the brief intervals around the crossings, we may apply the non-perturbative Landau-Zener theory presented above to the non-adiabatic evolution within the two-dimensional subspace spanned by the two crossing eigenstates. This yields the probability that the system will emerge, after the crossing, in the upper of the two crossing energy levels if it started in the lower, or vice versa.

When we start with all bosons occupying the $+$-level (as we have in Figure \ref{fig:EngineWork} and \ref{fig:VelDist}), the system is initially in the ground state of the eigenspectrum of (\ref{ReducedSchroedinger}). In the limit $N\gamma \ll\hbar k^{2}/M$, the only substantial opportunities for any bifurcation of the occupation amplitude distribution will actually be the $N$ crossings that lie along the lowest arc in the `web' shown in Supplementary Figure~\ref{fig:energylevels}. This is because in this limit we may treat  $\tilde{h}_{lm}$ as a perturbation, and in first order perturbation theory, $\tilde{h}_{lm}$ only couples $h_{m}$ eigenstates with $m$ differing by one. The crossings between these neighboring levels are precisely those along the lowest arc of the web; these are the points at which the two options are for one boson to descend from the $+$ to the $-$ mode (this is taking the lower branch of the crossing, and so continuing around the lowest arc of the web), or else for the boson distribution to remain unchanged (this is proceeding straight through the crossing, and so leaving the lowest arc). According to Landau-Zener theory as applied within the crossing subspace, the probability with which the system takes the upper branch of the $n^{\mathrm{th}}$ crossing, so that the transition from $n-1$ to $n$ bosons in the lower state \textit{does not} occur, is that stated in (\ref{Pn}), namely
\be\label{Pn2}
\mathcal{P}_{n} = \exp\left[-\frac{\pi\gamma^{2}}{2kg}n(N+1-n)\right] \;.
\ee
All crossings above the lowest arc in the Supplementary Figure~\ref{fig:energylevels} web are avoided only at higher order in perturbation theory, and hence in the limit $N\gamma \ll\hbar k^{2}/M$ are passed through in straight lines (\textit{i.e.} diabatically, with no change in boson distribution and hence no transfer of energy to the weight) with probability extremely close to one. In this limit, therefore, the chance that the engine will stall at any crossing may still be large or small, depending on the independent ratio $kg/\gamma^{2}$, but the chance that the engine will resume operation after once stalling is negligible.

When we transform the results of the reduced description (\ref{ReducedSchroedinger}) back to the full system, each crossing within the web of Supplementary Figure~\ref{fig:energylevels} corresponds to the weight hitting the critical velocity $v_c$ and undergoing a wave packet bifurcation, either jumping in velocity or not. Therefore, the probabilities (\ref{Pn2}) correspond directly to those of the probability distribution branches in Figure~\ref{fig:EngineWork} and \ref{fig:VelDist}. 

It is worth noting that although the lasting consequence of the crossing episode are simply the probabilities $P$ and $1-P$ of the two crossing levels, the brief non-adiabatic evolution that produces them is non-trivial, with probability generally oscillating back and forth between the two levels many times before settling down to the asymptotic results. This oscillating evolution shows up if one looks closely at Figure~\ref{fig:EngineWork}, as in Supplementary Figure~\ref{fig:MagCross}.

Generalizing these results beyond the simple regime of $N\gamma \ll\hbar k^{2}/M$ becomes complicated quickly, because the intervals of non-adiabatic evolution around the crossings can extend to encompass multiple crossings, and because the higher-order crossings also become important, leading to a complex flow of interfering quantum amplitudes through the entire web of $h_{n}$ levels. With numerical analysis in a wide range of regimes, however, even extending to the fully classical limit, we have confirmed that the micro-engine operates quite generally and robustly -- but always with some degree of inefficiency and uncertainty. We have also confirmed in the classical limit that the engine operates for a range of different nonlinear potentials. Our results therefore remain qualitatively valid within a much broader regime than explicitly discussed here.

\begin{figure}[h!]
\centering
\includegraphics[width=0.47\textwidth]{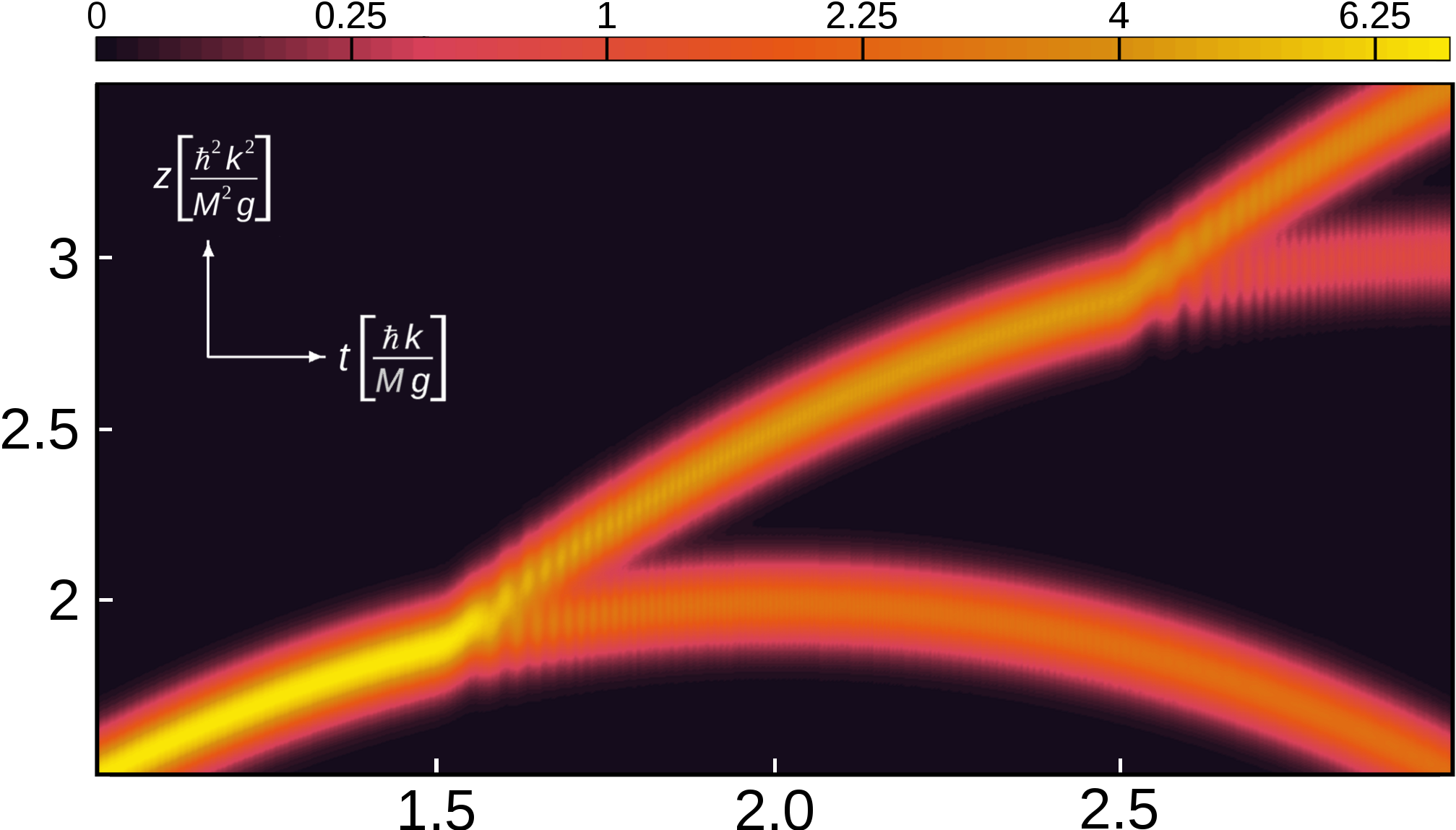}
\caption{Magnified view of the first probability bifurcation in Figure~\ref{fig:EngineWork}, showing the probability distribution $p(z,t)$ of the weight's height in the immediate vicinity of 
the first avoided crossing (`A') in the time-dependent reduced problem. The ripples visible on this scale show oscillation of probability between the two possibilities of stalling and working. For higher values of $\hbar k^{2}/(M\gamma)$, the non-adiabatic interval is so brief that the branches do not have time to separate appreciably within it, and these ripples become invisible. The color scaling is as in Figure~\ref{fig:EngineWork}.}
\label{fig:MagCross}
\end{figure}

\end{document}